\documentclass[reprint,aps,prl,showpacs,eqsecnum,twocolumn,superscriptaddress]{revtex4-1}
\usepackage{amsmath,amssymb,graphicx,color,ulem,multirow, inputenc}
\usepackage{hyperref,ulem,url}
\hypersetup{colorlinks=true}

\begin{document}

\title{Self-consistent picture of the mass ejection from a one second-long binary neutron star merger leaving a short-lived remnant in general-relativistic neutrino-radiation magnetohydrodynamic simulation} 

\author{Kenta Kiuchi}
\affiliation{Max Planck Institute for Gravitational Physics (Albert Einstein Institute), Am M\"{u}hlenberg, Potsdam-Golm, 14476, Germany}
\affiliation{Center for Gravitational Physics and Quantum Information, Yukawa Institute for Theoretical Physics, Kyoto University, Kyoto 606-8502, Japan}

\author{Sho Fujibayashi}
\affiliation{Max Planck Institute for Gravitational Physics (Albert Einstein Institute), Am M\"{u}hlenberg, Potsdam-Golm, 14476, Germany}

\author{Kota Hayashi}
\affiliation{Center for Gravitational Physics and Quantum Information, Yukawa Institute for Theoretical Physics, Kyoto University, Kyoto 606-8502, Japan}

\author{Koutarou Kyutoku}
\affiliation{Center for Gravitational Physics and Quantum Information, Yukawa Institute for Theoretical Physics, Kyoto University, Kyoto 606-8502, Japan}
\affiliation{Department of Physics, Kyoto University, Kyoto 606-8502, Japan}
\affiliation{Interdisciplinary Theoretical and Mathematical Science Program (iTHEMS), RIKEN, Wako, Saitama 351-0198, Japan}

\author{Yuichiro Sekiguchi}
\affiliation{Department of Physics, Toho University, Funabashi, Chiba 274-8510, Japan}
\affiliation{Center for Gravitational Physics and Quantum Information, Yukawa Institute for Theoretical Physics, Kyoto University, Kyoto 606-8502, Japan}

\author{Masaru Shibata}
\affiliation{Max Planck Institute for Gravitational Physics (Albert Einstein Institute), Am M\"{u}hlenberg, Potsdam-Golm, 14476, Germany}
\affiliation{Center for Gravitational Physics and Quantum Information, Yukawa Institute for Theoretical Physics, Kyoto University, Kyoto 606-8502, Japan}

\date{\today}

\begin{abstract}
We perform a general-relativistic neutrino-radiation magnetohydrodynamic simulation of a one second-long binary neutron star merger on Japanese supercomputer Fugaku using about $85$ million CPU hours with $20,736$ CPUs. We consider an asymmetric binary neutron star merger with masses of $1.2$ and $1.5M_\odot$ and a `soft' equation of state SFHo. It results in a short-lived remnant with the lifetime of $\approx 0.017$\,s, and subsequent massive torus formation with the mass of $\approx 0.05M_\odot$ after the remnant collapses to a black hole. For the first time, we find that after the dynamical mass ejection, which drives the fast tail and mildly relativistic components, the post-merger mass ejection from the massive torus takes place due to the magnetorotational instability-driven turbulent viscosity in a single simulation and the two ejecta components are seen in the distributions of the electron fraction and velocity with distinct features. 

\end{abstract}

\maketitle

{\it Intorduction}.--GW170817/AT~2017gfo/GRB~170817A heralded the beginning of multimessenger astrophysics~\cite{LIGOScientific:2017vwq,LIGOScientific:2017ync} and showed that binary neutron star (BNS) mergers are among the central observational targets  in such an era~\cite{LIGOScientific:2020aai}. By the observations of this merger event, the nuclear equation of state (EOS) of NS matter was constrained for the first time through the measurement of the tidal deformability in gravitational waves (GW)~\cite{LIGOScientific:2017vwq,LIGOScientific:2018hze,LIGOScientific:2018cki,De:2018uhw}. 
The elements heavier than iron were indicated to be synthesized via the rapid neutron capture process ($r$-process) nucleosynthesis in the neutron-rich matter ejected from the BNS merger~\cite{Metzger:2010,Lattimer:1974,Eichler:1989,Wanajo:2014wha}. This BNS merger drove a short gamma-ray burst (SGRB)~\cite{Goldstein:2017mmi,LIGOScientific:2017ync,Savchenko:2017ffs,Mooley:2018qfh}, which is a `smoking' gun for the merger hypothesis of the central engine~\cite{Paczynski:1986,Goodman:1985,Eichler:1989,Narayan:1992iy}.

By comparing the observational data with theoretical models of BNS mergers, we have managed to draw an overview about the merger process~\cite{Arcavi:2017xiz,Chornock:2017sdf,Coulter:2017wya,Cowperthwaite:2017dyu,Drout:2017ijr,Kasen:2017sxr,Kasliwal:2017ngb,Kilpatrick:2017mhz,McCully:2017lgx,Nicholl:2017ahq,Shappee:2017zly,Smartt:2017fuw,DES:2017kbs,Tanaka:2017qxj,Tanvir:2017pws,Goldstein:2017mmi,LIGOScientific:2017ync,Savchenko:2017ffs,Mooley:2018qfh,Kawaguchi:2018ptg,Perego:2017wtu,Radice:2017lry,Rossi:2019fnm,Milton:2018,Bauswein:2017vtn,LIGOScientific:2017pwl,Coughlin:2018miv,LIGOScientific:2018cki,Vieira:2022tnm,Fernandez:2018kax,Metzger:2021grk,Alexander:2017aly,Margutti:2017cjl,Fong:2017ekk,Margalit:2017dij,Margutti:2018xqd,Metzger:2018qfl,Hajela:2019mjy,Hajela:2021faz,Ishizaki:2021,Ioka:2017nzl,Hamidani:2019qyx,Matsumoto:2018mra,Shibata:2017xdx}. However, it does not mean that we have a consensus picture for this event. In particular, there is no consensus about the {\it detailed} mass ejection processes~\cite{Shibata:2017xdx,Kasen:2017sxr,Kiuchi:2019lls,Radice:2017lry,Metzger:2018qfl,Fujibayashi:2017puw,Perego:2017wtu,Waxman:2017sqv,Kawaguchi:2018ptg,Breschi:2021tbm,Nedora:2019jhl,Nedora:2020hxc,Siegel:2018,Villar:2017wcc}, although neutron-rich matter was likely to be ejected during the BNS merger process~\cite{Shibata:2019} because the electromagnetic signals associated with the so-called kilonova/macronova emissions were detected~\cite{Metzger:2010,Li:1998bw,Kulkarni:2005}. Also, there is no theoretical consensus about how the BNS merger drove the relativistic jet in this event~\cite{Milton:2018,Rezzolla:2017aly,Mosta:2020hlh,Fernandez:2018kax,Christie:2019lim}. All these situations encourage us to build an accurate theoretical model of BNS mergers.

Numerical relativity is the chosen tool to explore BNS mergers from inspiral to post-merger phases theoretically. Recent studies for the BNS merger remnants have indicated that it is mandatory to perform a simulation in which the neutrino-radiation transfer and mangnetohydrodynamics are taken into account {\it at least} for $O(1)$\,s to explore the entire mass ejection and relativistic jet launching processes~\cite{Fujibayashi:2022ftg,Fujibayashi:2020qda,Fujibayashi:2020jfr,Fujibayashi:2020dvr,Christie:2019lim,Fernandez:2018kax,Shibata:2021bbj,Mosta:2020hlh} . This simulation timescale is also required by the time lag of $\approx 1.7$\,s observed between GW170817 and GRB~170817A~\cite{LIGOScientific:2017vwq,LIGOScientific:2017ync}. 

However, all the previous works suffer from the limitation such as the short simulation time of $\approx 0.1$\,s~\cite{Radice:2018a,Zappa:2022,Miller:2019dpt,Radice:2017lry,Radice:2018a,Bernuzzi:2016pie,Foucart:2022kon,Milton:2018,Raithel:2022nab,Most:2021ktk,Vigano:2020ouc}, non-self-consistent initial conditions of the merger remnants constructed by equilibrium configuration of massive tori around a black hole (BH)~\cite{Christie:2019lim,Siegel:2018,Fernandez:2018kax,Miller:2019dpt,Fujibayashi:2020jfr}, or a phenomenological prescription to model turbulent viscosity induced by the magnetorotational instability (MRI~\cite{Balbus-Hawley:1991})~\cite{Fujibayashi:2022ftg,Fujibayashi:2020qda,Fujibayashi:2020jfr,Fujibayashi:2020dvr,Radice:2017zta,Foucart:2022kon,Radice:2018a,Metzger:2008wj}. 
To obtain a self-consistent picture for the merger and post-merger evolution of BNSs, it is necessary to perform a  neutrino-radiation magnetohydrodynamics simulation in full general relativity at least for one second. 

We tackle this problem using Japanese supercomputer Fugaku. Specifically, we focus on a BNS merger leaving a short-lived remnant as suggested to be the case for GW170817 by the absence of a strong radio emission~\cite{Shibata:2017xdx,Metzger:2018qfl,Margalit:2017dij}, and also to be typical cases by the universality of the $r$-process elemental abundance~\cite{Fujibayashi:2022ftg}. We delineate a comprehensive picture of the BNS merger from the inspiral to post-merger phases. In this Letter, we report the mass, the chemical property, and the velocity profile for both the dynamical and post-merger ejecta derived in a single simulation.

{\it Numerics, model, and grid steup}.--
We employ a neutrino-radiation magnetohydrodynamics code in numerical relativity~\cite{Kiuchi:2022} for the BNS merger simulation. We solve Einstein's equation by the BSSN-puncture formulation together with the Z4c constraint propagation prescription~\cite{Shibata:1995,Baumgarte:1998te,Baker:2005vv,Campanelli:2005dd,Hilditch:2012fp} using 4th-order accurate finite difference in time and space. We also employ the HLLD Riemann solver~\cite{MUB:2009} together with the constrained transport solver~\cite{Gardiner:2007nc} to evolve the equations for relativistic magnetohydrodynamics. The neutrino-radiation field is solved with the truncated moment formalism~\cite{Shibata:2011kx} together with a gray general relativistic leakage prescription to handle the neutrino cooling~\cite{Sekiguchi:2010ep,Sekiguchi:2012uc}. The neutrino heating is also taken into account~\cite{Fujibayashi:2017xsz}.

The NS is modeled with the SFHo EOS~\cite{Steiner:2012rk}, for which the maximum mass of spherical NSs is $2.06M_\odot$. We here consider an asymmetric binary with mass of $1.2$ and $1.5M_\odot$. The LORENE liberary~\cite{LORENE} is employed to construct a quasi-equilibrium configuration of the BNS with the initial orbital frequency of $Gm_0\Omega_0/c^3=0.025$ where $m_0=2.7M_\odot$ is the total mass and an orbital eccentricity of $O(10^{-3})$~\cite{Kyutoku:2014yba}. We extend the original SFHo EOS table to the low-density and temperature with the Helmholtz EOS~\cite{Timmes}.
The floor values for the density and temperature are $0.166~{\rm g~cm^{-3}}$ and $10^{-3}$~MeV, respectively.

To cover a wide dynamic range of the problem, we employ a fixed mesh refinement (FMR) in the Cartesian coordinates~\cite{Kiuchi:2022} with the {\it reflux} prescription and the divergence-free- and magnetic-flux-preserving prolongation for the magnetic field~\cite{Balsara:2001}. The number of the FMR domain is set to be $13$. The grid resolution in a coarser FMR domain is twice as large as that in an adjacent finer domain; $\Delta x_{l-1}=2\Delta x_l$ with $l=2,\cdots, 13$. The size of the finest domain is $L_{13}\in[-37.875~{\rm km},37.875~{\rm km}]$ with the grid spacing $\Delta x_{13}=150$~m in the $x$, $y$, and $z$ directions. The $z=0$ plane is the orbital plane (see Supplement Material for the detailed setup).
We also perform a simulation with $\Delta x_{13}=200$~m to check the convergence.

A poloidal magnetic field is initialized by the vector potential prescription~\cite{Kiuchi:2017zzg,Aguilera-Miret:2021fre}:
\begin{align}
A_{(\varphi)} = A \max(P-2\times 10^{-4}P_{\rm max},0)^2,
\end{align}
where $P$ and $P_{\rm max}$ are the pressure and its maximum value, respectively. $A$ is set to be such that the initial maximum magnetic-field strength is $10^{15}$\,G. Although the initial magnetic field is stronger than those observed in binary pulsars~\cite{Lorimer:2008se}, the Kelvin-Helmholtz instability is likely to amplify the magnetic-field strength to the magnetar level in a very short time at the onset of merger in reality~\cite{Kiuchi:2014,Kiuchi:2017zzg,Kiuchi:2015,Rasio:1999,Price:2006,Aguilera-Miret:2021fre}, and thus, the numerical results should not depend strongly on the initial field strength as far as the grid resolution is sufficiently high.

\begin{figure*}[t]
 	 \includegraphics[width=0.99\linewidth]{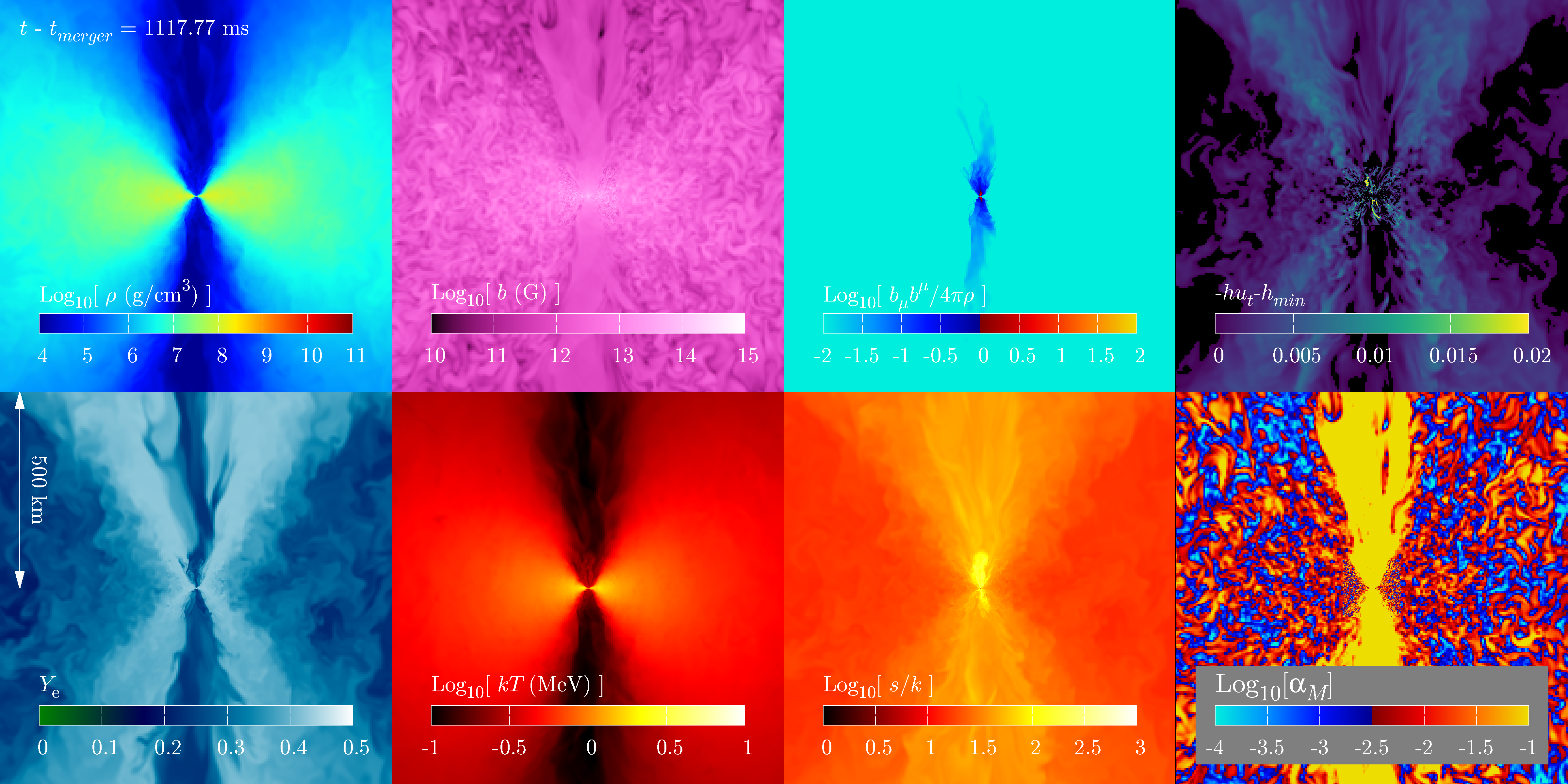}
 	 \caption{Profiles for rest-mass density (top-left), magnetic-field strength (top-second from left), magnetization parameter (top-second from right), unboundedness defined by the Bernoulli criterion (top-right), electron fraction (bottom-left), temperature (bottom-second from left), entropy per baryon (bottom-second from right), and Shakura-Sunyaev $\alpha_\mathrm{M}$ parameter (bottom-right) on the $y-y_\mathrm{AH}=0$ plane at $t-t_\mathrm{merger}\approx 1.1$~s. See also the movie: \url{http://www2.yukawa.kyoto-u.ac.jp/~kenta.kiuchi/anime/FUGAKU/out_SFHo_12_15.mp4}. 
      }\label{fig:2D_plot}
\end{figure*}

{\it Overview}.--Figure~\ref{fig:2D_plot} plots the profiles of the rest-mass density $\rho$, the magnetic-field strength, the magnetization parameter defined by $\sigma_B = b^\mu b_\mu/\rho c^2$ with the magnetic field in the fluid rest frame $b^\mu$, unboundedness defined by the Bernoulli criterion, the electron fraction $Y_e$, the temperature, the entropy per baryon, and Shakura-Sunyaev $\alpha_\mathrm{M}$ parameter~\cite{Shakura:1972te} due to the Maxwell stress defined by $\alpha_\mathrm{M} \equiv \langle -b^{(r)}b_{(\varphi)}/ P \rangle$ on the $y-y_\mathrm{AH}=0$ plane at $t-t_\mathrm{merger}\approx 1.1$~s. $y_\mathrm{AH}$ denotes the location of the puncture point in the $y$ direction. We define the merger time $t_\mathrm{merger}$ at which the GW amplitude becomes maximum. $\langle \cdot \rangle$ denotes the time average over the time interval of $1$~ms. 
The following describes how the system evolves toward the final state (see also Supplement Material). 

\begin{figure*}[t]
	 \includegraphics[scale=0.25]{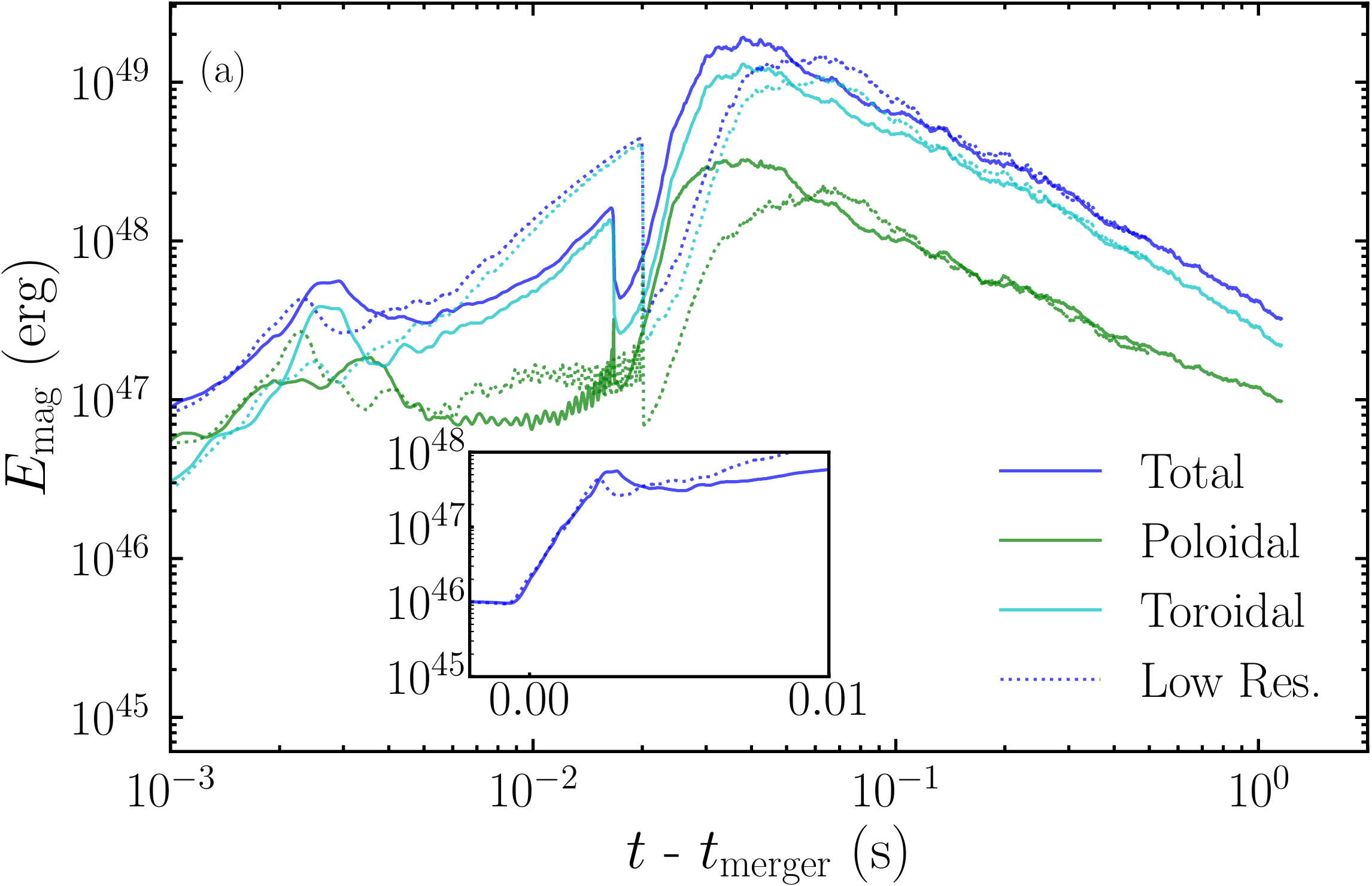}
 	 \includegraphics[scale=0.25]{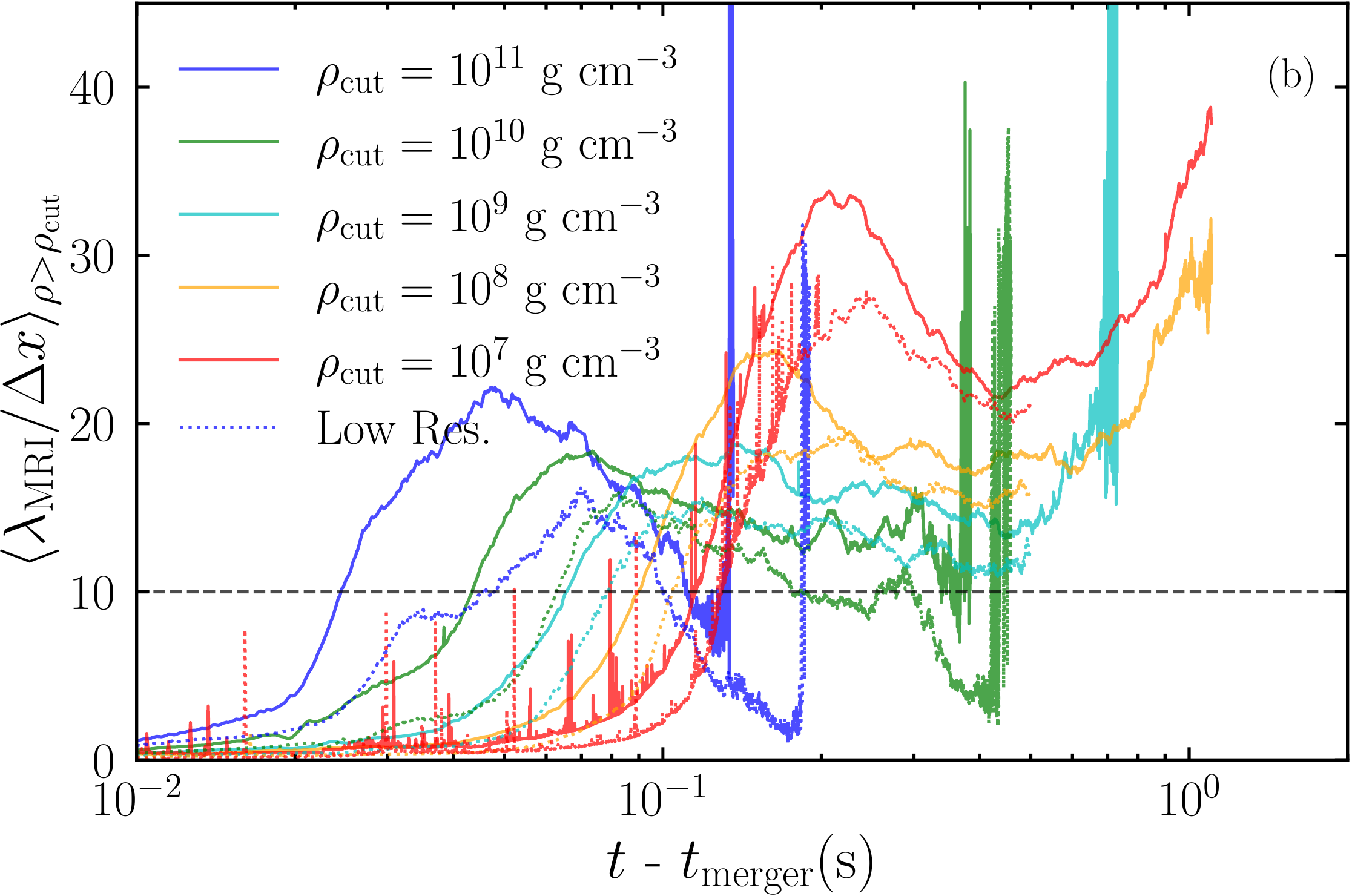}
	 \includegraphics[scale=0.25]{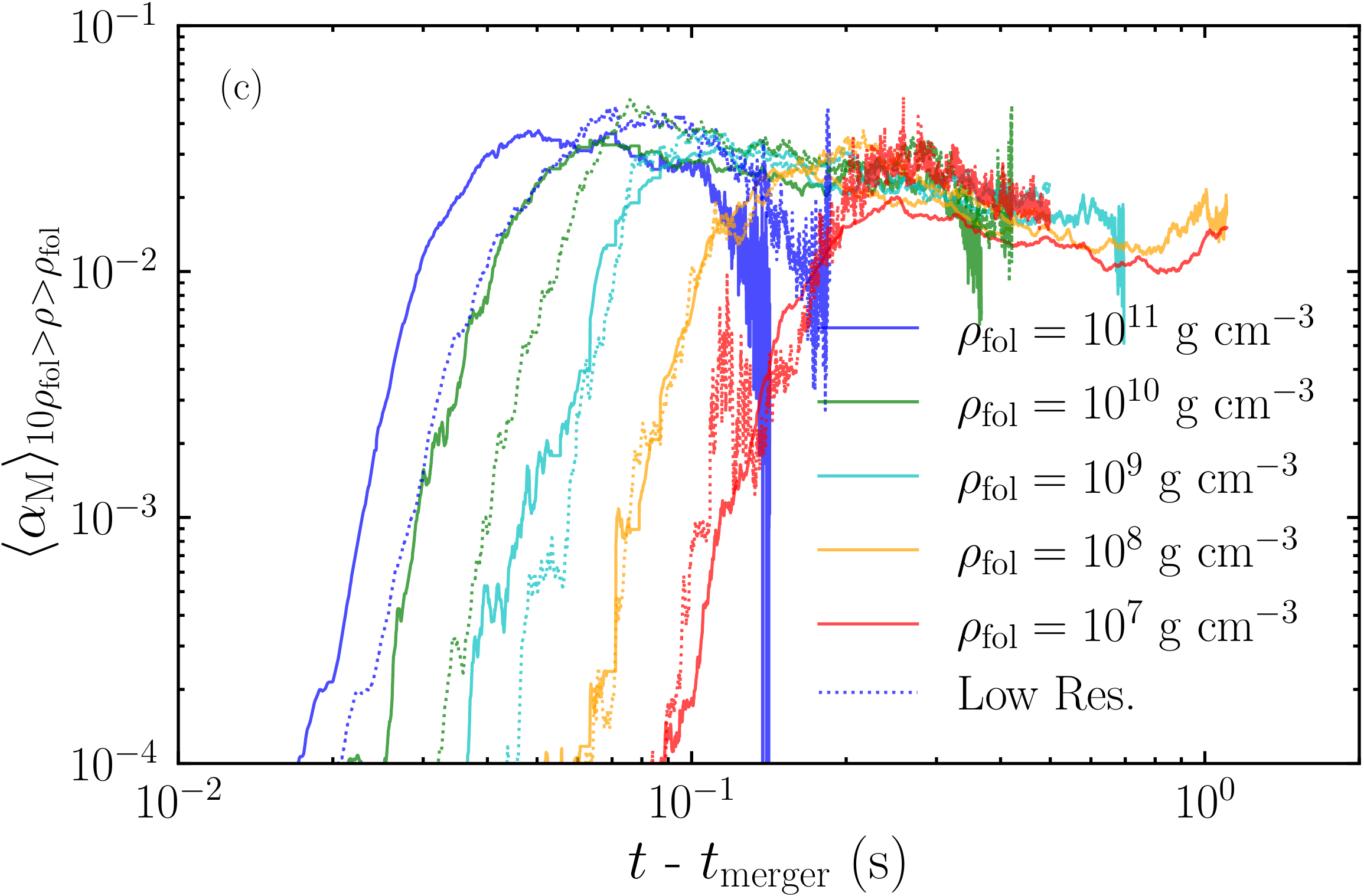}
 	 \includegraphics[scale=0.25]{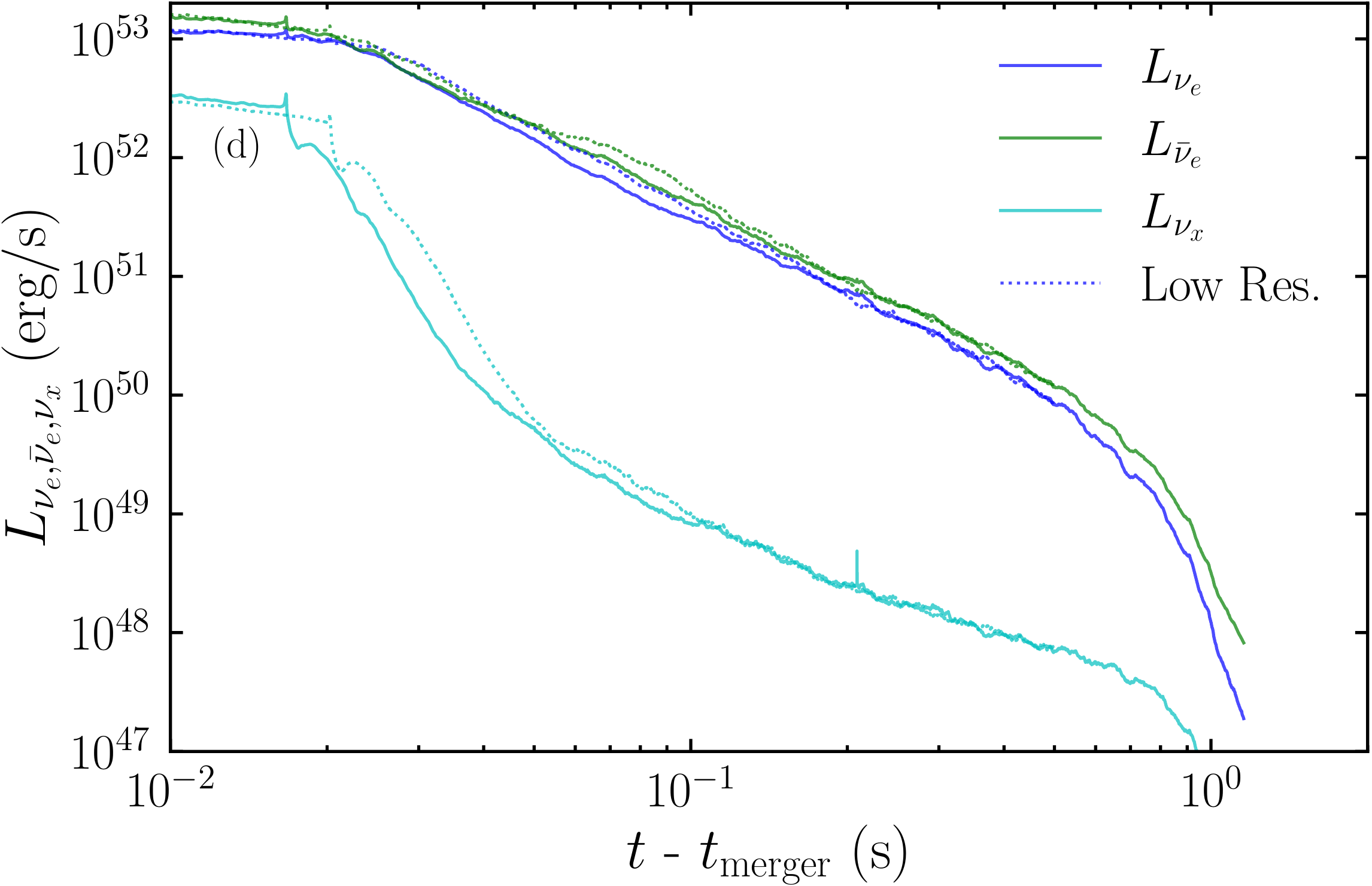}
 	 \includegraphics[scale=0.25]{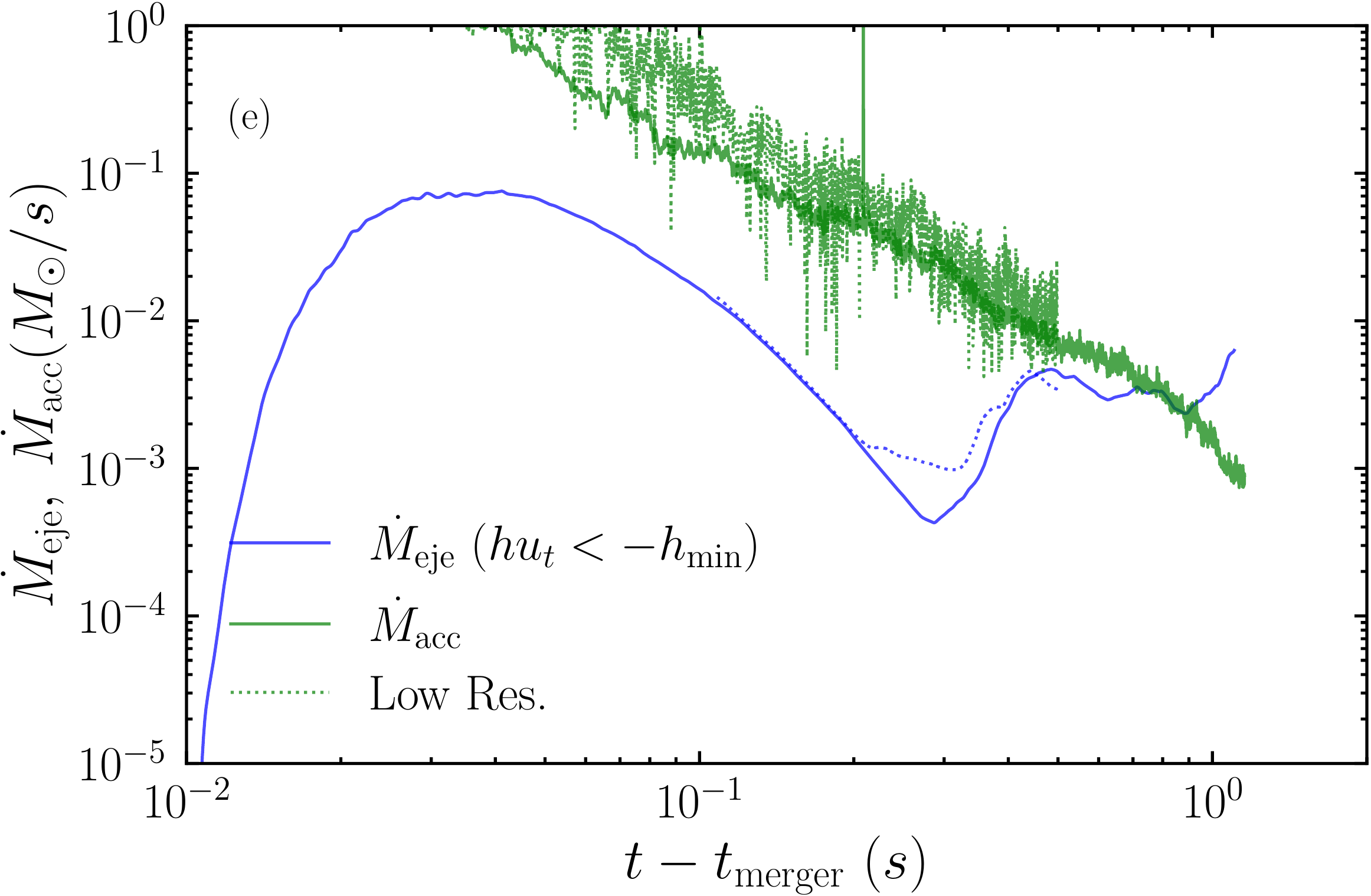}
     \includegraphics[scale=0.25]{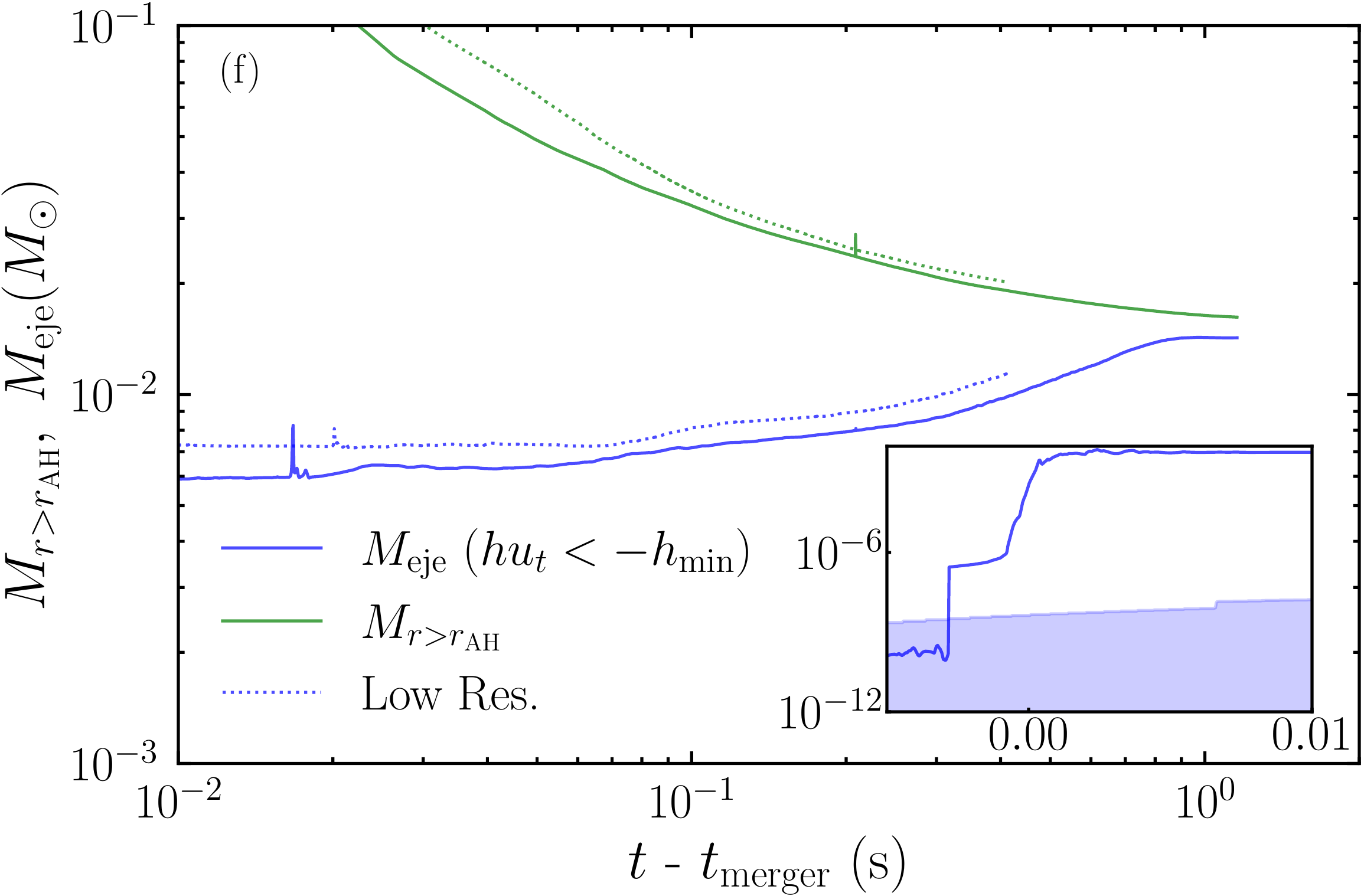}
 	 \caption{
 	   Time evolution of various quantities. (a) Electromagnetic energy. The green and cyan curves denote the poloidal and toroidal components, respectively. The inset shows the total electromagnetic energy evolution at the merger. (b) Volume-averaged MRI quality factor for selected cutoff rest-mass densities. (c) Shakura-Sunyaev $\alpha_\mathrm{M}$ parameter for selected foliation rest-mass densities. (d) Neutrino luminosity for the electron $(\nu_e)$, anti-electron $(\bar{\nu}_e)$, and heavy species $(\nu_x)$. (e) Mass ejection rate measured on the sphere of $r_\mathrm{sphere}\approx 3,000$~km (blue) and mass accretion rate onto the BH (green). (f) Baryonic mass outside the horizon (green) and ejecta (blue). The inset shows the ejecta around the merger time, and the color-shaded region denotes the baryonic mass conservation error. The BH formation time is $t-t_\mathrm{merger}\approx 0.017$~s.  In panels (b) and (c), the blow-up and the rapid decrease behavior, respectively, indicate the disappearance of the fluid elements with the corresponding rest-mass density. In all the figures, the dotted curves denote the result with $\Delta x_{13}=200$~m.
         }\label{fig:Mdot}
\end{figure*}

{\it Magnetic-field amplification and MRI dynamo}.--In the present simulation, a short-lived hypermassive NS (HMNS) is formed after the  five-orbit inspiral and merger. The magnetic field is amplified after the merger (see Fig.~\ref{fig:Mdot}~(a)) due to the Kelvin-Helmholtz instability, subsequent magnetic winding, and non-axisymmetric MRI~\cite{Kiuchi:2014,Aguilera-Miret:2021fre}. The HMNS collapses to a BH at $t-t_\mathrm{merger}\approx 0.017$~s~\cite{Sekiguchi:2015dma,Kiuchi:2022,Fujibayashi:2022ftg}. 

During the HMNS phase, the non-axisymmetric density structure of the star exerts a gravitational torque on the fluid elements. As a result, the angular momentum is transported outward. Thus, a massive torus is formed at the BH formation. The mass of the torus is $\approx 0.05M_\odot$ at $t-t_\mathrm{merger} \approx 0.04$~s at which the electromagnetic energy saturates. The mass and dimensionless spin of the BH are $\approx 2.55M_\odot$ and $\approx 0.65$, respectively. The magnetic field is amplified inside the torus due to the magnetic winding and subsequent radial motion of the fluid elements resulting from the enhanced magnetic pressure just after the BH formation as shown in Fig.~\ref{fig:Mdot}~(a)~\cite{Kiuchi:2014}. The fastest growing mode of the axisymetric MRI~\cite{Balbus-Hawley:1998} starts to be resolved for $t-t_\mathrm{merger} \gtrsim 0.02$~s in the high-density region of the torus. The electromagnetic energy saturates at $\approx 1$~\% of the internal energy of the torus. Subsequently, the electromagnetic energy decreases with time due to mass accretion.

To quantify how the fastest growing mode of the MRI is well resolved, we define the volume-averaged MRI quality factor with a density cutoff $\rho_\mathrm{cut}$ by
\begin{align}
Q_{\mathrm{MRI},\rho_\mathrm{cut}} \equiv \left\langle \frac{\lambda_\mathrm{MRI}}{\Delta x} \right\rangle_{\rho_\mathrm{cut}} \equiv \frac{\int_{\rho \ge \rho_\mathrm{cut}}\lambda_\mathrm{MRI} d^3x}{\Delta x\int_{\rho \ge \rho_\mathrm{cut}} d^3x},
\end{align}
where $\lambda_\mathrm{MRI}=2\pi b^z/(\sqrt{\rho h+b^\mu b_\mu}\Omega)$ with $h$ the specific enthalpy and $\Omega$ the angular velocity. Figure~\ref{fig:Mdot}~(b) plots $Q_{\mathrm{MRI},\rho_\mathrm{cut}}$ with $\rho_\mathrm{cut}=10^{7,8,9,10,11}~{\rm g~cm^{-3}}$, and it clearly shows that the MRI is well resolved in a bulk region of the torus for $t-t_\mathrm{merger}\gtrsim 0.1$~s. 

Once the MRI sets in, the MRI-driven turbulence is developed. As a result, effective turbulent viscosity is enhanced. The turbulent viscosity facilitates the angular momentum transport and heats up the matter due to the viscous heating. As a result, the torus expands outward, and the mass accretion onto the BH is facilitated (see Fig.~\ref{fig:2D_plot} and Fig.~\ref{fig:Mdot}~(e)). The bottom-right panel of Fig.~\ref{fig:2D_plot} shows that the estimated  Shakura-Sunyaev $\alpha_\mathrm{M}$ parameter has the spatial distribution with $\sim 10^{-3}$--$10^{-1}$ in the bulk region of the torus, and $\sim 10^{-1}$ in the vicinity of the torus surface. Figure~\ref{fig:Mdot}~(c) plots the evolution of $\alpha_\mathrm{M}$ foliated in terms of the rest-mass density $\rho_\mathrm{fol}$:
\begin{align}
\left \langle \alpha_\mathrm{M} \right \rangle_{10\rho_\mathrm{fol} \ge \rho \ge \rho_\mathrm{fol}} \equiv \frac{\int_{10\rho_\mathrm{fol} \ge \rho \ge \rho_\mathrm{fol}}\alpha_\mathrm{M} d^3x}{\int_{10 \rho_\mathrm{fol} \ge \rho \ge \rho_\mathrm{fol}} d^3x}.
\end{align}
It shows that the MRI-driven turbulent viscosity is generated once the MRI is resolved, and the saturation value varies in the range of $\approx 10^{-2}$--$3\times 10^{-2}$ depending on $\rho_\mathrm{fol}$.

\begin{figure}[t]
 	 \includegraphics[scale=0.38]{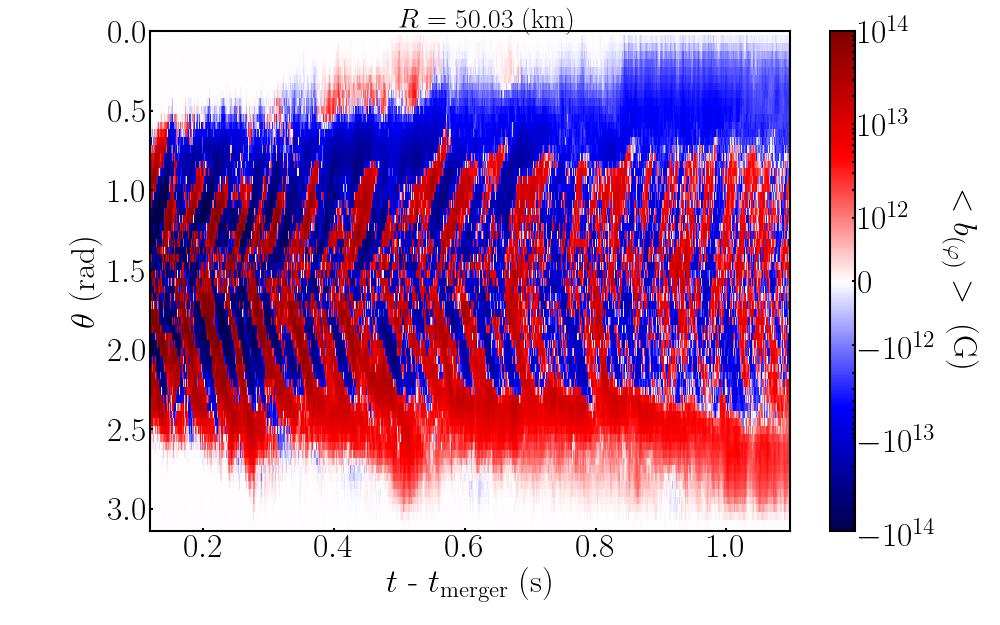}
 	 \caption{Butterfly diagram for the azimuthally-averaged toroidal magnetic field on a sphere with a radius of $R\approx 50$~km. 
         }\label{fig:Butterfly_Diagram}
\end{figure}

The MRI-driven turbulence is sustained by the MRI dynamo~\cite{Balbus-Hawley:1998}. To show the MRI-dynamo activity, we generate a butterfly diagram of the azimuthally-averaged toroidal magnetic field measured on a meridional line with a radius of $\approx 50~{\rm km}$ in Fig.~\ref{fig:Butterfly_Diagram}. It clearly shows the sign-flip pattern which lasts until the end of the simulation.

As the torus expands, the temperature decreases by the adiabatic cooling. As a result of this, the neutrino luminosity decreases with time as shown in Fig.~\ref{fig:Mdot}~(d) because the neutrino emissivity is approximately proportional to $T^6$~\cite{Fuller:1985}. 

{\it Mass ejection}.--Due to the angular momentum transport and heating facilitated by the MRI-driven turbulent viscosity, a part of the torus is ejected as the post-merger ejecta. Figure~\ref{fig:Mdot}~(e) plots the mass ejection rate measured on a sphere with a radius of $\approx 3,000$~km. We identify the ejecta as the fluid elements which satisfy the Bernoulli criterion with the positive radial velocity~\cite{Fujibayashi:2022ftg}. The plot shows the rapid rise of the ejection rate around $t-t_\mathrm{merger}\approx 0.01$~s. The ejection rate peaks at $t-t_\mathrm{merger}\approx 0.03$--$0.04$~s, and it decreases with time. This component corresponds to the dynamical ejecta composed of the fast tail with the terminal velocity $v_\infty \approx 0.96{\rm c}$~\cite{Fujibayashi:2022ftg,Hotokezaka:2018,Metzger:2014yda} and the mildly relativistic component with the average velocity of $v_\infty \approx 0.25{\rm c}$~\cite{Sekiguchi:2016bjd}. At $t-t_\mathrm{merger}\approx 0.3$~s, a new component emerges, which corresponds to the post-merger ejecta~\cite{Fujibayashi:2022ftg,Fernandez:2013tya,Just:2014fka,Just:2021cls}. We find that the major driving force of this component is {\it not} the Lorentz force, but the MRI-driven turbulent viscosity because the plasma beta is much larger than unity when the ejecta is launched (see also Fig.~2 in Supplement Material). 

The top-right panel of Fig.~\ref{fig:2D_plot} shows the morphology of this post-merger ejecta. It is initially driven along the torus surface. Subsequently, the outer part of the torus is ejected to the equatorial direction (see Fig.~1 in Supplement Material). The turbulent viscosity-driven post-merger ejection still lasts for $t-t_\mathrm{merger} \gtrsim 1$~s. 

The mass ejection rate exceeds the mass accretion rate for $t-t_\mathrm{merger} \gtrsim 0.8~{\rm s}$, and the neutrino luminosity steeply decreases for $t-t_\mathrm{merger} \gtrsim 0.7~{\rm s}$, which indicates the neutrino cooling becomes inefficient and most of the turbulent viscous heating energy can be used for the torus expansion~\cite{Fujibayashi:2022ftg,Fujibayashi:2020qda,Just:2021cls,Just:2014fka}. Table~\ref{tab:ejecta} shows the mass of the dynamical and post-merger ejecta, as well as the torus mass at $t-t_\mathrm{merger}\approx 1.1$~s. Figure~\ref{fig:Mdot}~(f) also plots the evolution of the baryonic mass outside the apparent horizon and the ejecta mass. Because the mass ejection rate and accretion rate at $t-t_\mathrm{merger}\approx 1.1$\,s is $\approx 6\times10^{-3}M_\odot/{\rm s}$ and $\approx 10^{-3}M_\odot/{\rm s}$, respectively, we expect that the post-merger mass ejection will continue for another  $O(0.1$--$1)$~s. We confirm all the quantities relevant to the post-merger mass ejection enter the convergent regime with respect to the grid resolution (see Fig.~\ref{fig:Mdot}).

\begin{figure*}[t]
	 \includegraphics[scale=0.23]{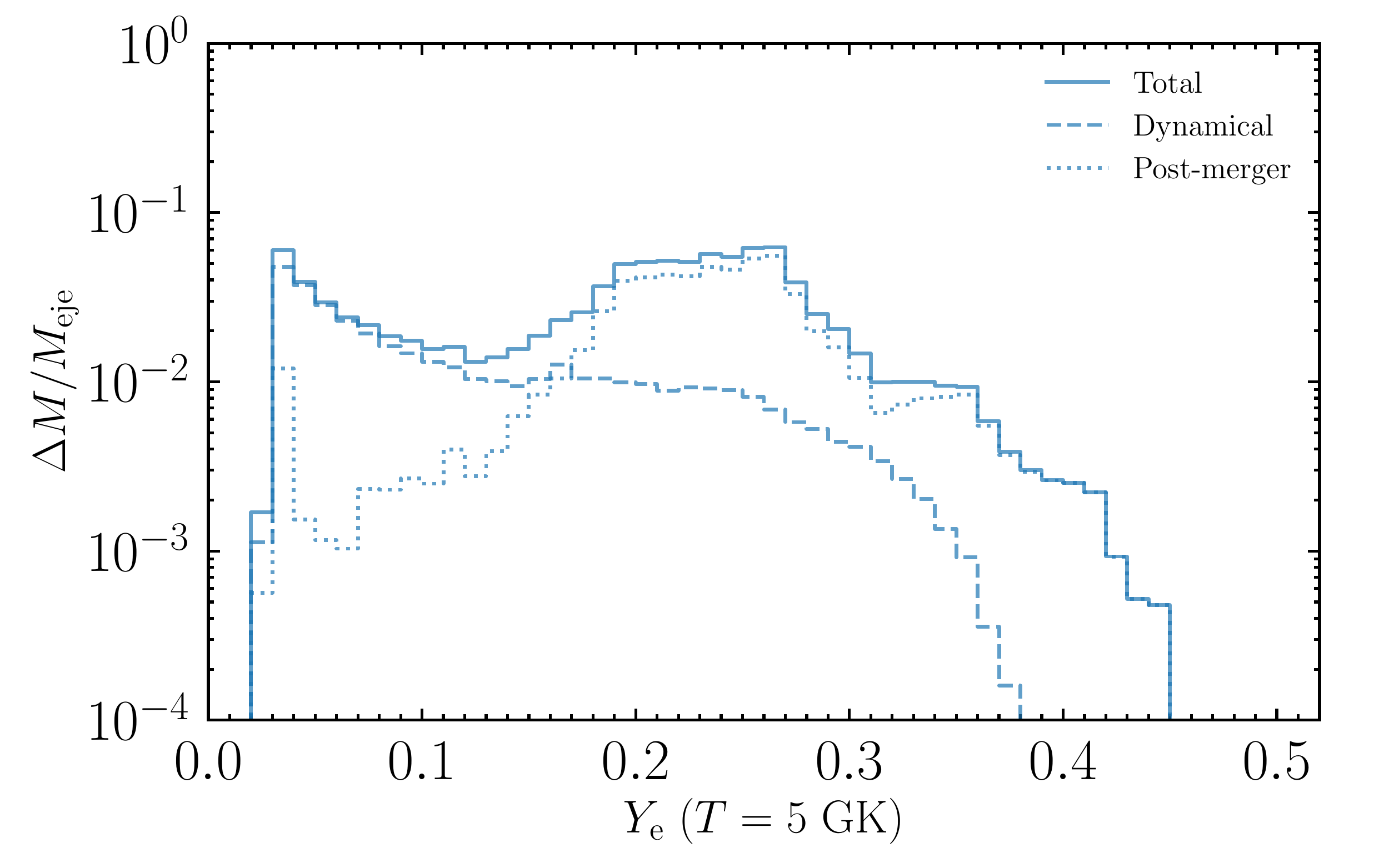}
	 \includegraphics[scale=0.23]{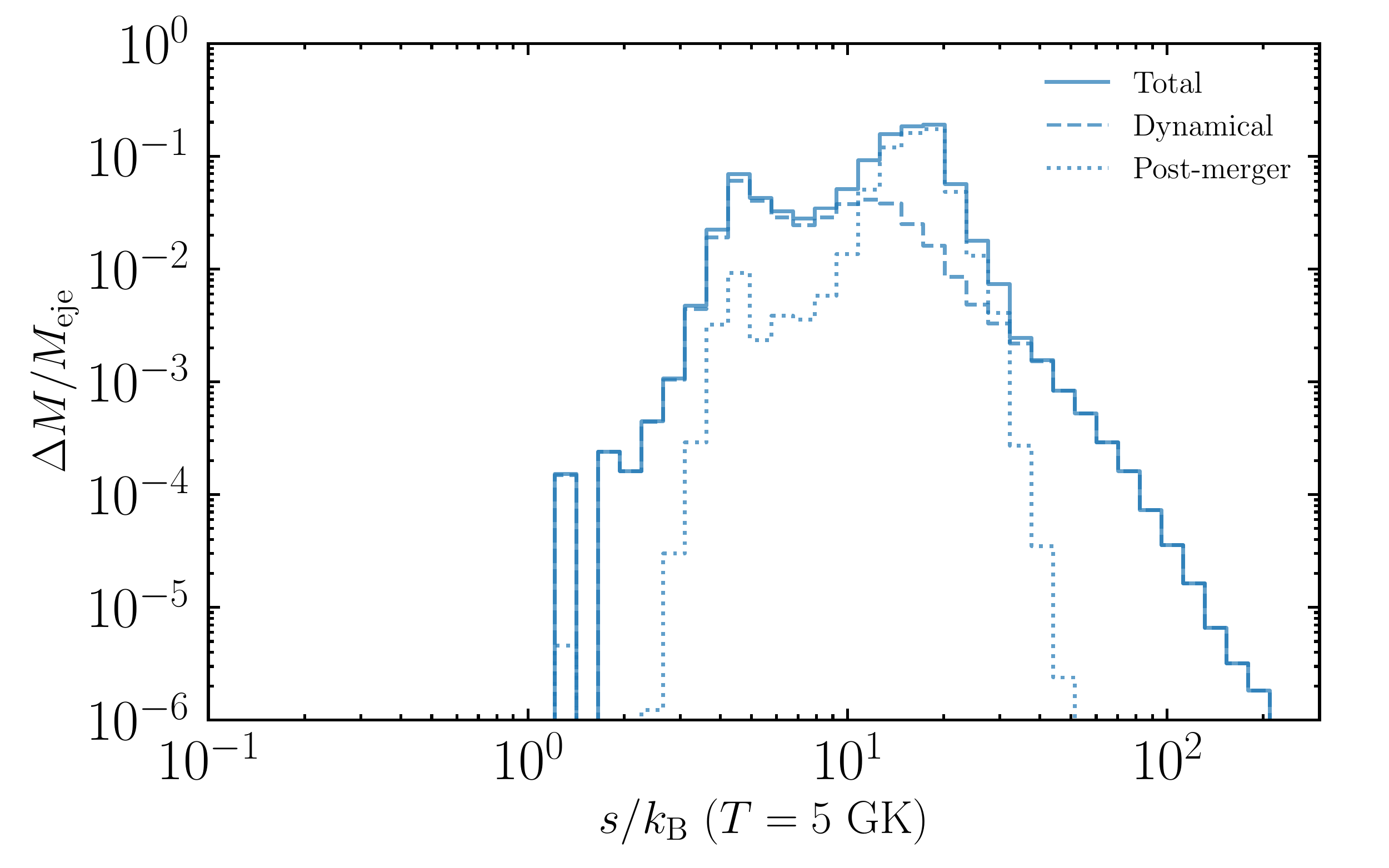}
	 \includegraphics[scale=0.23]{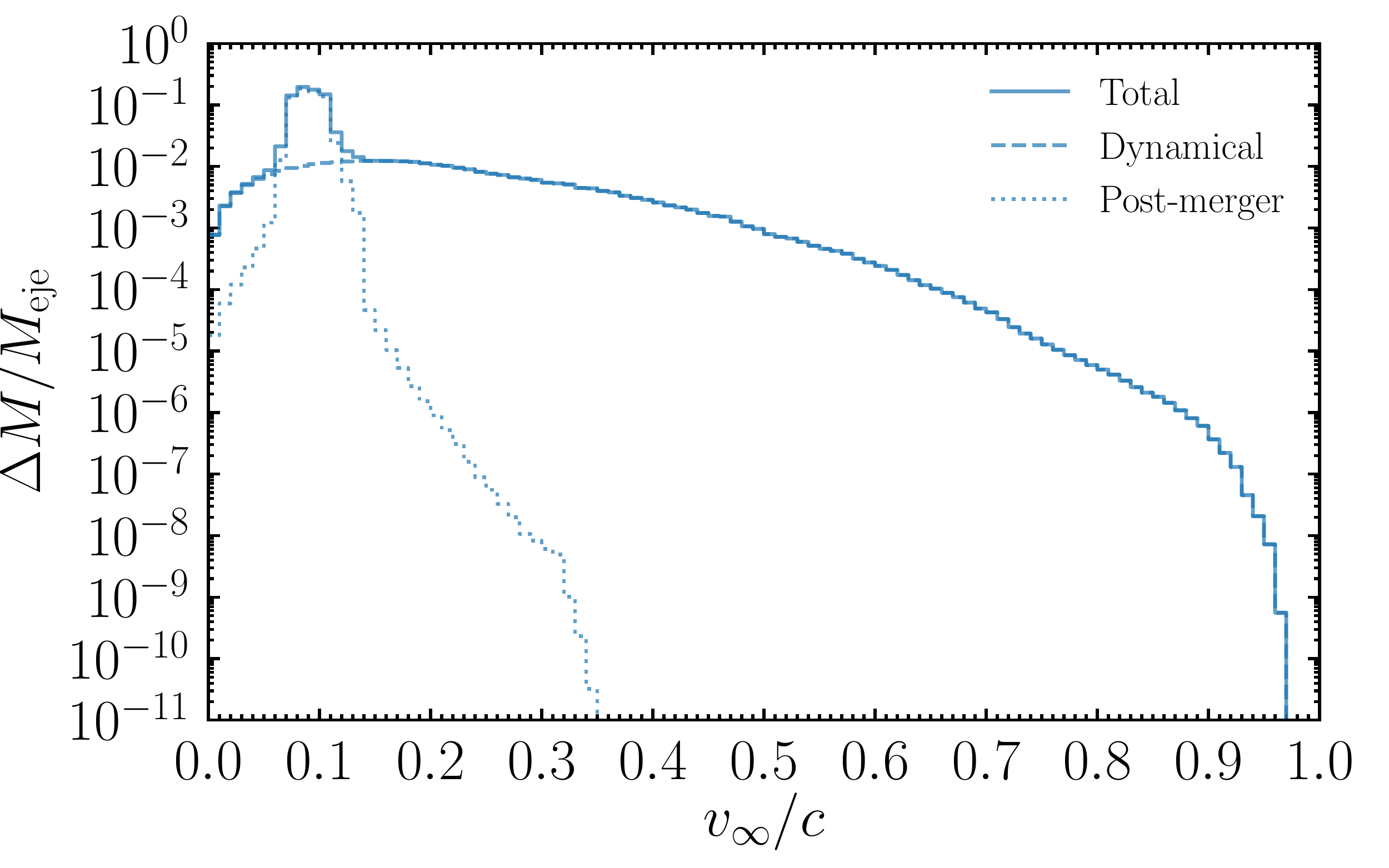}
 	 \caption{
 	 Mass histogram of the ejecta as a function of the electron fraction (left), the entropy per baryon (center), and the terminal velocity (right) at $t-t_\mathrm{merger}\approx 1.1~{\rm s}$ calculated by the tracer particle. The solid, dashed, and dotted curves denote the profiles for the total, dynamical, and post-merger ejecta, respectively.
         }\label{fig:ejecta_profile}
\end{figure*}

Figure~\ref{fig:ejecta_profile} plots the mass histogram of the ejecta as a function of the electron fraction, entropy per baryon, and terminal velocity at $t-t_\mathrm{merger}\approx 1.1~{\rm s}$ for the total, dynamical, and post-merger components, respectively. The electron fraction profile has two distinct peaks at $Y_e\approx 0.03$ and $Y_e \approx 0.26$--$0.27$. This profile is similar to our previous result based on the viscous hydrodynamics for the post-merger ejecta~\cite{Fujibayashi:2022ftg} although a slight difference is also found between two models. The agreement of $Y_{e,\mathrm{peak}}\approx0.26$--$0.27$ to that found in the one-zone model of the accretion disk around the merger remnant with $\alpha_M=0.03$ and $M_\mathrm{torus}=0.1(0.01)M_\odot$ in Ref.~\cite{Metzger:2008wj} is reasonable although the latter model predicts slightly higher value of $Y_{e,\mathrm{peak}} \approx 0.30(0.34)$. The fraction of the initial disk mass that remains at the end of the simulation is $\approx 4\%$ in our model and $15$--$21\%$ in their models. The entropy profile also shows the multimodal structure, which has peaks at $s/k_\mathrm{B}\approx 3$, $10$, and $20$. The low-$Y_e$ component corresponds to two peaks at $s/k_\mathrm{B}\approx 3$ and $10$. They represent the dynamical ejecta, and the $s/k_\mathrm{B}\approx 3$ and $10$ peaks correspond respectively to the tidal- and shock-driven components~\cite{Sekiguchi:2015dma,Sekiguchi:2016bjd,Bauswein:2013yna,Hotokezaka:2012ze}. The high-$Y_e$ component corresponds to $s/k_\mathrm{B}\approx 20$, and this represents turbulent viscosity-driven post-merger ejecta. The peak value of $Y_e$ for the post-merger ejecta is determined when the weak interaction freezes out~\cite{Fujibayashi:2022ftg}. The terminal velocity profile also shows that the dynamical ejecta extends up to $\approx 0.96$\,c and the post-merger ejecta exhibits a peak around $\approx 0.08$--$0.1$\,c.

{\it Summary}.--We performed the neutrino-radiation magnetohydrodynamic simulation of the BNS merger in numerical relativity for one second, focusing on the short-lived HMNS formation in an asymmetric binary merger, which results in the massive torus formation after collapsing to a BH. We confirmed the development of the MRI dynamo inside the torus, which produces a fully turbulent state. The resultant turbulent viscosity facilitated the angular momentum transport and generated a quasi-steady heating source. The merger remnant composed of the BH and a massive torus was evolved up to $\approx 1.1$~s after the merger. 

We found that the dynamical ejecta was driven by the tidal force and shock heating at the merger and the subsequent post-merger mass ejection was driven primarily by the MRI-driven turbulent viscosity from the torus after the neutrino cooling becomes inefficient in the single simulation. The ejecta contains the dynamical component with $Y_{e,\mathrm{peak}}\approx 0.03$ and $s_\mathrm{peak}/k_\mathrm{B}\approx 3$ and $\approx 10$, and the post-merger component with $Y_{e,\mathrm{peak}}\approx 0.26$--$0.27$ and $s_\mathrm{peak}/k_\mathrm{B}\approx 20$. Because the peak value of $Y_e$ for the post-merger ejecta is close to the critical value of $\approx 0.25$ of the $r$-process nucleosynthesis for the lanthanide elements~\cite{Tanaka:2018}, it could result in the efficient lanthanide element production. 

Our model suggests that the SFHo EOS may be disfavored as the nuclear equation of state because simple kilonova light-curve modelings of AT 2017gfo often require $M_\mathrm{eje} \approx 0.05M_\odot$~\cite{Hotokezaka:2020}. The detailed $r$-process nucleosynthesis calculation~\cite{Wanajo:2014wha} and kilonova light-curve modeling~\cite{Kawaguchi:2022bub} are on going. 

Our simulation unifies pictures of the mass ejection from the BNS merger leaving a short-lived HMNS. We did not find the Poynting-flux dominated outflow to the polar direction within the simulation time perhaps because of the only moderately rapid BH spin, the ram pressure due to the fallback motion of the dynamical ejecta, the shortness of the simulation time and/or a spurious BH spin down caused by the insufficient grid resolution (see Supplement Material). With a longer-term run, in which the rest-mass density in the polar region would decrease, it might be possible to find the launch of the strong Poynting flux. A simulation with a timescale much longer than  1\,s and finer grid resolution is a remaining issue. It is also necessary to implement a sophisticated neutrino-radiation transfer scheme to precisely predict the properties of the post-merger ejecta~\cite{Just:2015fda}.

\begin{table}
\caption{Mass of the dynamical ejecta, the turbulent-driven post-merger ejecta, and the torus at $t-t_\mathrm{merger}\approx 1.1$~s.}\label{tab:ejecta}
\begin{tabular}{ccc}
\hline\hline
$M_\mathrm{eje,dyn}~(M_\odot)$ & $M_\mathrm{eje,post}~(M_\odot)$ & $M_\mathrm{torus}~(M_\odot)$\\
\hline\hline
$6 \times 10^{-3}$ & $8 \times 10^{-3}$ & $2 \times 10^{-3}$ \\
\hline
\end{tabular}
\end{table}

{\it Acknowledgments}.--This work used computational resources of the supercomputer Fugaku provided by RIKEN through the HPCI System Research Project (Project ID: hp220174). The simulation was also performed on Sakura, Cobra, and Raven clusters at the Max Planck Computing and Data Facility and on the Cray XC50 at CfCA of the National Astronomical Observatory of Japan. This work was in part supported by the Grant-in-Aid for Scientific Research (grant Nos. 20H00158, 22K03617, and 23H04900) of Japan MEXT/JSPS. Kiuchi thanks to the Computational Relativistic Astrophysics members in AEI for a stimulating discussion. 


\bibliography{reference}

\end{document}